\documentclass[aps,prl,twocolumn,showpacs,amsmath,amssymb]{revtex4-1}
\usepackage{graphicx}  
\usepackage{subfigure}
\usepackage{epstopdf}
\usepackage{dcolumn}   
\usepackage{bm}        
\usepackage{amssymb}   
\usepackage{amsmath}
\usepackage{mathrsfs}
\usepackage{color}
\usepackage[textsize=scriptsize]{todonotes}
\usepackage{lineno}
\usepackage{ulem}
\usepackage{color}
\usepackage{overpic}

\newcommand{\mev}{\mathrm{MeV}}

\newcommand{\mevcc}{\mathrm{MeV}/c^2}

\newcommand{\gevcc}{\mathrm{GeV}/c^2}

\begin{document}

\widetext

\title{$\Sigma^{+}$ and $\bar{\Sigma}^-$ polarization in the $J/\psi$ and $\psi(3686)$ decays}
\author{
\begin{small}
\begin{center}
M.~Ablikim$^{1}$, M.~N.~Achasov$^{10,d}$, P.~Adlarson$^{64}$, S. ~Ahmed$^{15}$, M.~Albrecht$^{4}$, A.~Amoroso$^{63A,63C}$, Q.~An$^{60,48}$, ~Anita$^{21}$, Y.~Bai$^{47}$, O.~Bakina$^{29}$
% ...
\\
\vspace{0.2cm}
(BESIII Collaboration)\\
\vspace{0.2cm} {\it
$^{1}$ Institute of High Energy Physics, Beijing 100049, People's Republic of China\\
% ...
$^{4}$ Bochum Ruhr-University, D-44780 Bochum, Germany\\
% ...
$^{10}$ G.I. Budker Institute of Nuclear Physics SB RAS (BINP), Novosibirsk 630090, Russia\\
% ...
$^{15}$ Helmholtz Institute Mainz, Johann-Joachim-Becher-Weg 45, D-55099 Mainz, Germany\\
% ...
$^{21}$ Indian Institute of Technology Madras, Chennai 600036, India\\
% ...
$^{29}$ Joint Institute for Nuclear Research, 141980 Dubna, Moscow region, Russia\\
% ...
$^{47}$ Southeast University, Nanjing 211100, People's Republic of China\\
$^{48}$ State Key Laboratory of Particle Detection and Electronics, Beijing 100049, Hefei 230026, People's Republic of China\\
% ...
$^{60}$ University of Science and Technology of China, Hefei 230026, People's Republic of China\\
% ...
$^{63}$ (A)University of Turin, I-10125, Turin, Italy; (B)University of Eastern Piedmont, I-15121, Alessandria, Italy; (C)INFN, I-10125, Turin, Italy\\
$^{64}$ Uppsala University, Box 516, SE-75120 Uppsala, Sweden\\
% ...
$^{d}$ Also at the Novosibirsk State University, Novosibirsk, 630090, Russia\\
% ...
}\end{center}
\vspace{0.4cm}
\end{small}
}

\begin{abstract}
From $1310.6\times10^{6}$ $J/\psi$ and $448.1\times10^{6}$ $\psi(3686)$ events collected with the BESIII experiment, we report the first observation of 
$\Sigma^{+}$ and $\bar{\Sigma}^{-}$ spin polarization in $e^+e^-\rightarrow J/\psi (\psi(3686)) \rightarrow \Sigma^{+} \bar{\Sigma}^{-}$ decays. The relative phases of the form factors $\Delta\Phi$ have been measured to be $(-15.5\pm0.7\pm0.5)^{\circ}$ and $(21.7\pm4.0\pm0.8)^{\circ}$ with $J/\psi$ and $\psi(3686)$ data, respectively.
The non-zero value of $\Delta\Phi$ allows for a direct and simultaneous measurement of the decay asymmetry parameters of $\Sigma^{+}\rightarrow p \pi^{0}~(\alpha_0 = -0.998\pm0.037\pm0.009)$ and $\bar{\Sigma}^{-}\rightarrow \bar{p} \pi^{0}~(\bar{\alpha}_0 = 0.990\pm0.037\pm0.011)$, the latter value being determined for the first time. The average decay asymmetry, $(\alpha_{0} - \bar{\alpha}_{0})/2$,  is calculated to be $-0.994\pm0.004\pm0.002$. 
The CP asymmetry $A_{\rm CP,\Sigma} = (\alpha_0 + \bar{\alpha}_0)/(\alpha_0 - \bar{\alpha}_0) = -0.004\pm0.037\pm0.010$ is extracted for the first time, and is found to be consistent with CP conservation.
\end{abstract}

\maketitle

Hyperons are ideal probes for studying the strong interaction in the transition region between the non-perturbative and perturbative QCD regimes. In addition, two-body hyperon weak decays play an important role in the study of symmetry properties in particle physics. 

Historically, these decays were used to establish parity violation~\cite{Lee:1957he}. Current research on this type of decays focuses on the search for CP-violation in the baryon sector. 
The polarization of spin 1/2 hyperons can be determined in two-body weak decays due to the
self-analyzing nature of these decay processes. 
The $\Sigma^{+}$ polarization vector $\mathbf{P}_{\Sigma^{+}}$ can be determined from the $\Sigma^+\to p\pi^0$ decay using the angular distribution of the daughter proton, as $dN/d\Omega = \frac{1}{4\pi}( 1 + \alpha_{0}\mathbf{P}_{\Sigma^{+}} \cdot \mathbf{\hat{p}})$. 
Here, $\mathbf{\hat{p}}$ is the unit vector along the proton momentum in the $\Sigma^{+}$ rest frame and $\alpha_{0}$ is defined as the decay asymmetry parameter for the $\Sigma^{+}\to p \pi^{0}$ decay. Correspondingly, the 
decay asymmetry parameter for $\Sigma^{-}\to \bar{p} \pi^{0}$ is denoted $\bar{\alpha}_{0}$. 
The parameters $\alpha_{0}$ and $\bar{\alpha}_{0}$ are CP-odd so that $A_{\rm CP,\Sigma}=(\alpha_{0}+\bar{\alpha}_{0})/(\alpha_{0}-\bar{\alpha}_{0})$ can be used to test CP-symmetry ~\cite{Okubo:1958zza, Pais:1959zza}.
A non-zero value of $A_{\rm CP, \Sigma}$ would indicate CP-violation. An average decay asymmetry parameter $\alpha_{0} = -0.980^{+0.017}_{-0.013}$~\cite{Tanabashi:2018oca} was extracted from $\pi^{+} p \rightarrow \Sigma^{+} K^{+}$ experiments nearly fifty years ago~\cite{Harris:1970kq, Bellamy:1972fa, Lipman:1973mz} while $\bar{\alpha}_{0}$ has not been measured before.
The Standard Model theoretical prediction for the level of CP-violation is $A_{\rm CP, \Sigma}\sim 3.6\times10^{-6}$~\cite{Tandean:2002vy}. 
In general, CP-violation in the baryonic sector is relatively poorly known~\cite{Aaij:2016cla}.
It has thus been noted in Ref.~\cite{Bigi:2017eni} that it is of high importance to improve the sensitivity regarding CP-violation in as many baryonic decay modes as possible in order to investigate the consistency with the Standard Model Cabibbo-Kobayashi-Maskawa mechanism.

BESIII provides a unique environment to study both hyperon production and decay properties in electron-positron annihilation to $\Sigma^{+}$  $\bar{\Sigma}^{-}$ pairs via the intermediate $J/\psi$ and $\psi'$ (denoting the $\psi(3686)$ throughout this letter) resonances ~\cite{Bigi:2017eni}. In this quantum entangled system, the decay parameters of the two baryons are correlated which allows a controlled and precise test of CP-symmetry.
Recently, the first case of hyperon polarization in electron-positron annihilation was found for $\Lambda$ hyperons in the $J/\psi\to\Lambda\bar{\Lambda}$ decay by the BESIII collaboration~\cite{Ablikim:2018zay}.

\begin{figure}[b]
\includegraphics[width=0.45\textwidth]{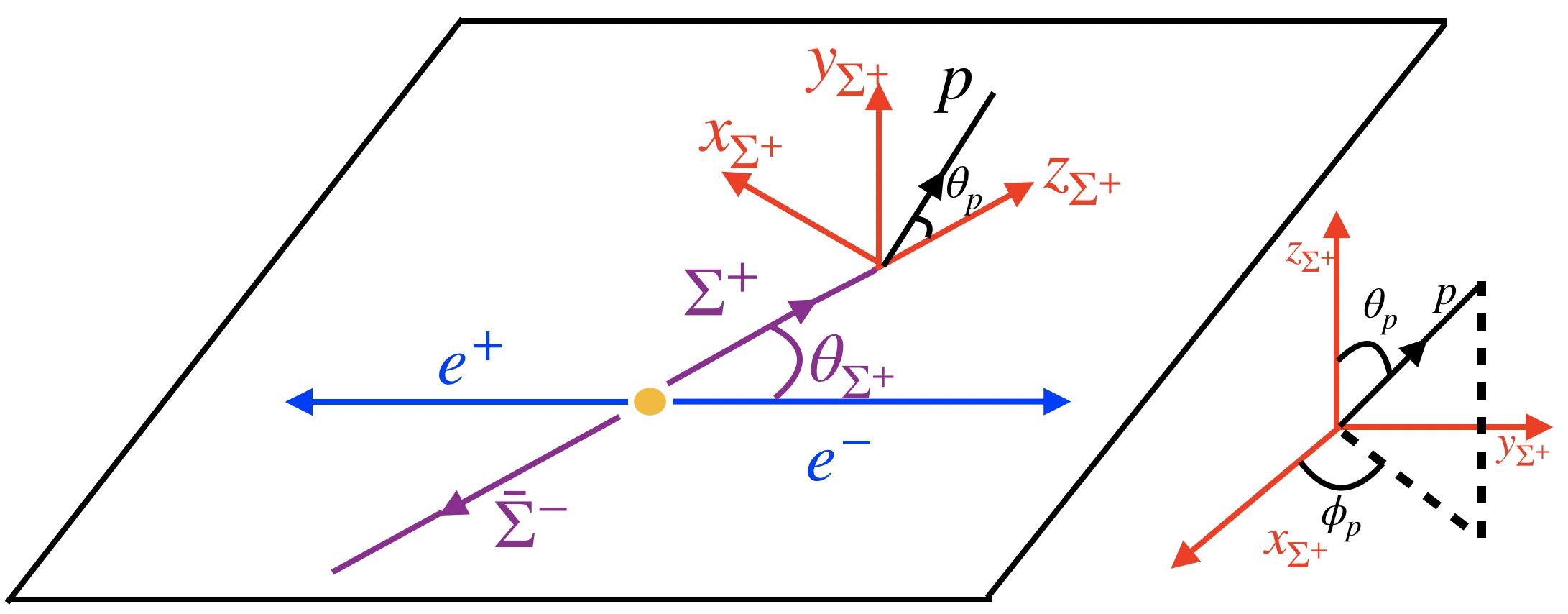}
\caption{(Color online) Definition of the coordinate system used to describe the $J/\psi \rightarrow \Sigma^{+} \bar{\Sigma}^{-}$ and $\psi' \rightarrow \Sigma^{+} \bar{\Sigma}^{-}$ process. The $\Sigma^{+}$ particle is emitted along the $z_{\Sigma^{+}}$ axis direction, and the $\bar{\Sigma}^{-}$ in the opposite direction. $y_{\Sigma^{+}}$ axis is perpendicular to the plane of $\Sigma^{+}$ and $e^{-}$, and $x_{\Sigma^{+}}$ axis is defined by right-hand coordinate system. The $\Sigma^{+}$ decay product, proton, is measured in this coordinate.}
\label{fig:helicity_frame}
\end{figure}

The $e^{+}e^{-}\to \Psi \to \Sigma^{+}\bar{\Sigma}^{-}$ ($\Psi$ here denotes either the $J/\psi$ or the $\psi'$)  production process is described by the psionic electric and magnetic form factors, $G_{E}^{\Psi}$ and $G_{M}^{\Psi}$ ~\cite{Faldt:2017kgy}. These two psionic form factors are formally equivalent to the $\Sigma$ electric and magnetic form factors~\cite{Dubnickova:1992ii,Gakh:2005hh,Czyz:2007wi,Faldt:2013gka,Faldt:2016qee}. The two form factors can be described by two real parameters $\alpha_{\Psi}$ and $\Delta\Phi$, which correspond to the angular decay asymmetry and the relative phase between the form factors, respectively. The observable $\Delta\Phi$ is related to the spin-polarization of the produced $\Sigma^{+} \bar{\Sigma}^{-}$ pair. In singly weak decays, if the relative phase is non-zero $\Delta\Phi \neq 0$, the $\Sigma$ polarization is perpendicular to the production plane and depends on the angle between the $\Sigma^{+}$ and electron ($e^{-}$) beam in the reaction center-of-mass frame (CM) $\theta_{\Sigma^{+}}$, as shown in Fig.~\ref{fig:helicity_frame}. It is then possible to make a simultaneous and direct measurement of $\alpha_{0}$ and $\bar{\alpha}_{0}$ and hence also a test on CP-symmetry.

The first branching fraction measurement of $J/\psi \to \Sigma^{+} \bar{\Sigma}^{-}$ was reported by the BES collaboration~\cite{Ablikim:2008tj} while $\psi' \rightarrow \Sigma^{+} \bar{\Sigma}^{-}$ was studied with CLEO data~\cite{Pedlar:2005px, Dobbs:2014ifa, Dobbs:2017hyd}. However, so far no measurement of $\alpha_{\Psi}$ and $\Delta\Phi$ exists. 

The full differential cross-section of the production and decay process $e^{+}e^{-}\rightarrow \Psi \rightarrow \Sigma^{+}(\to p\pi^{0}) \bar{\Sigma}^{-}(\to\bar{p}\pi^{0})$ is described with five observables $\boldsymbol{\xi}= ( \theta_{\Sigma^{+}}, \theta_{p}, \phi_{p}, \theta_{\bar{p}}, \phi_{\bar{p}})$~\cite{Faldt:2017kgy}. 
Here $\theta_{p}, \phi_{p}$ and $\theta_{\bar{p}}, \phi_{\bar{p}}$ are the polar and azimuthal angles of the proton and anti-proton measured in the rest frames of their respective mother  particles. As seen from the basis vector definitions in Fig.~\ref{fig:helicity_frame}, the $z$-axis is taken along the $\Sigma^+$ momentum $\textbf{p}_{\Sigma^{+}} = - \textbf{p}_{\bar{\Sigma}^{-}} = \textbf{p}$ in the CM system. The $y$-axis is taken as the normal to the scattering plane, $\textbf{k}_{e^{-}} \times \textbf{p}_{\Sigma^{+}}$, where $\textbf{k}_{e^{-}} = - \textbf{k}_{e^{+}} = \textbf{k}$ is the electron beam momentum in the CM system. Forming a right-handed coordinate system, the basis vectors are
\begin{eqnarray}
	{\textbf{x}}_{\Sigma^{+}} &=& \frac{1}{\sin{\theta_{\Sigma^{+}}}}(\hat{\textbf{k}} \times \hat{\textbf{p}})\times \hat{\textbf{p}} , \ \
	{\textbf{y}}_{\Sigma^{+}} = \frac{1}{\sin{\theta_{\Sigma^{+}}}}(\hat{\textbf{k}} \times \hat{\textbf{p}} ), \ \  \nonumber \\
	{\textbf{z}}_{\Sigma^{+}} &=& \hat{\textbf{p}}.
\end{eqnarray}

The differential cross-section is given as $d\sigma \propto  {\cal{W}}({\boldsymbol{\xi}}) d{\boldsymbol{\xi}}$, where ${\cal{W}}({\boldsymbol{\xi}})$ is
\begin{widetext} 
\begin{equation}
\begin{split}
	{\cal{W}}({\boldsymbol{\xi}})=	 &{\cal{T}}_0({\boldsymbol{\xi}})+{{\alpha_{\psi}}}{\cal{T}}_5({\boldsymbol{\xi}})\\
	+&{{\alpha_{0}}{\bar{\alpha}_{0}}}\left({\cal{T}}_1({\boldsymbol{\xi}})
+\sqrt{1-\alpha_{\psi}^2}\cos({{\Delta\Phi}}){\cal{T}}_2({\boldsymbol{\xi}})
+{{\alpha_{\psi}}}{\cal{T}}_6({\boldsymbol{\xi}})\right)\\
+&\sqrt{1-\alpha_{\psi}^2}\sin({{\Delta\Phi}})
\left({\alpha_{0}}{\cal{T}}_3({\boldsymbol{\xi}})
+\bar{{\alpha}}_{0}{\cal{T}}_4({\boldsymbol{\xi}})\right).\label{eq:anglW}
\end{split}
\end{equation}
\end{widetext}
and ${\cal{T}}_{i}, (i = 0, 1...6)$ are angular functions dependent on  $\boldsymbol{\xi}$ which are described in detail in Ref.~\cite{Faldt:2017kgy}.
Equation ~(\ref{eq:anglW}) contains three types of terms: those depending on the $\Sigma^{+}$ scattering angle (${\cal{T}}_{0}+\alpha_{\psi}{\cal{T}}_5$), the three spin correlations (terms multiplied by $\alpha_{0}\bar{\alpha}_{0}$) 
and the separate polarization terms. 
The free parameters to be determined in a fit of the joint angular distribution ${\cal{W}}({\boldsymbol{\xi}})$ to the data are $\alpha_{\Psi}$, $\Delta\Phi$, $\alpha_{0}$ and $\bar{\alpha}_{0}$.
If $\Delta\Phi \neq 0$, all parameters can be determined simultaneously, and $A_{\rm CP}$ can be evaluated directly.

In this letter, we present a study of the $J/\psi \to \Sigma^{+}\bar{\Sigma}^{-}$ and $\psi'\to \Sigma^{+}\bar{\Sigma}^{-}$ decays. In an analysis of the angular distributions of the $\Sigma^{+}$ ($\bar{\Sigma}^{-}$) baryons and their daughter particles, the spin polarization and decay asymmetry parameters of $\Sigma^{+}$ and $\bar{\Sigma}^{-}$ are measured for the first time.

The analysis is based on  $1310.6\times10^{6}$ $J/\psi$ and $448.1\times10^{6}$ $\psi'$ events collected with the BESIII detector. The BESIII detector is a magnetic
spectrometer~\cite{Ablikim:2009aa} located at the Beijing Electron
Positron Collider (BEPCII)~\cite{Yu:IPAC2016-TUYA01}. The
cylindrical core of the BESIII detector consists of a helium-based
 multilayer drift chamber (MDC) for measuring
the momenta and specific ionization energy loss ($dE/dx$) of
charged particles, a plastic scintillator time-of-flight
system (TOF) which contributes to charged particle identification (PID), and a CsI(Tl) electromagnetic calorimeter (EMC),
which are all enclosed in a superconducting solenoidal magnet
providing a 1.0~T (0.9~T during $J\/\psi$ data taking in
2012) magnetic field. The solenoid is supported by an
octagonal flux-return yoke with resistive plate counter muon
identifier modules interleaved with steel. The acceptance of
charged particles and photons is 93\% over $4\pi$ solid angle. The
charged-particle momentum resolution at $1~{\rm GeV}/c$ is
$0.5\%$, and the $dE/dx$ resolution is $6\%$ for the electrons
from Bhabha scattering. The EMC measures photon energies with a
resolution of $2.5\%$ ($5\%$) at $1$~GeV in the barrel (end cap)
region. The time resolution of the TOF barrel part is 68~ps, while
that of the end cap part is 110~ps.

Candidate events for the process $\Psi \rightarrow \Sigma^{+} \bar{\Sigma}^{-}$, with subsequent  $\Sigma^{+}$($\bar{\Sigma}^{-})\rightarrow p\pi^{0}$($\bar{p}\pi^{0}$) and $\pi^{0} \rightarrow \gamma \gamma$ decays, have to have two good charged tracks with opposite charges and at least four photons. Good charged tracks are required to be within the acceptance of the MDC, $|\cos\theta|<0.93$.
For each track, the point of closest approach to the interaction point must be within 2~cm in the plane perpendicular to the beam direction and within $\pm$10~cm along the beam direction. The two good charged tracks need to be identified as proton and anti-proton by the PID system, requiring that the likelihood for a proton assignment is larger than alternative hypotheses, $\mathcal{L}(p) > \mathcal{L}(\pi)$ and $\mathcal{L}(p) > \mathcal{L}(K)$. Here, $\mathcal{L}(h)$ $(h = \pi, K, p)$  is a likelihood for the different final state hadron hypotheses determined from the specific energy loss in the MDC and the time-of-flight measurement.

Photon candidates are reconstructed from isolated showers in the EMC. Each photon candidate is required to have a minimum energy of 25~$\mev$ in the EMC barrel region ($|\cos\theta|<0.8$) or 50~$\mev$ in the endcap region ($0.86<|\cos\theta|<0.92$). To improve the reconstruction efficiency and the energy resolution, the energy deposited in the nearby TOF counters is included in the photon reconstruction. In order to further suppress electronic noise and energy deposition unrelated to the signal event, it is required that the time-difference between an EMC signal and the reconstructed event start time is within an interval of $700$~ns. Good $\pi^{0}$ candidates are selected as those photon pairs whose invariant mass is satisfying
$(m_{\pi^0}-60~\mevcc) <M_{\gamma \gamma} < (m_{\pi^0}+40~\mevcc)$, where $m_{\pi^0}$ is the nominal
mass of the $\pi^0$ meson~\cite{Tanabashi:2018oca}. An asymmetric mass window is used for the $\pi^0$ reconstruction as the photon energy deposited in the EMC has a tail on the low energy side.
In addition, a one-constraint (1C) kinematic fit is performed for the photon pairs, constraining the invariant mass to the nominal $\pi^0$ mass. The $\chi^2_{\rm 1C}$ of the kinematic fit is required to be less than 25.
The number of good $\pi^0$ candidates is required to be larger than one.
To further remove potential background events and improve the mass resolution, a four-constraint (4C) kinematic fit is performed, constraining the total reconstructed four
momentum to that of the initial state. A requirement on the quality of the 4C kinematic fit of $\chi^2_{\rm 4C} < 100$ is imposed. If the number of $\pi^0$ candidates in an event is greater than two, the $p\bar{p}\gamma\gamma\gamma\gamma$ combination with the lowest $\chi^{2}_{\rm 4C}$ is selected as the final event candidate.
After kinematic fitting, the $\Sigma^{+}$ and $\bar{\Sigma}^{-}$ candidates are built from the proton-, anti-proton- and neutral pion-candidates. Here, the combination that minimizes $\sigma_m = \sqrt{(M_{p\pi^{0}} -m_{\Sigma^{+}})^2 + (M_{\bar{p}\pi^{0}} -
m_{\bar{\Sigma}^{-}})^2}$ is chosen in order to allocate the neutral pions to the two baryon decays.
For the $\psi' \rightarrow \Sigma^{+}\bar{\Sigma}^{-}$ decay, an additional invariant mass requirement is imposed on the proton-antiproton pair, $|M_{p\bar{p}} - 3.1~\gevcc| > 0.05 ~\gevcc$, to remove background events of the decay $\psi' \to \pi^0 \pi^0  J/\psi$ with $J/\psi \rightarrow p \bar{p}$.

To investigate possible background processes in the final data sample, inclusive Monte Carlo (MC) samples of $1.2 \times 10^9$~$J/\psi$ and $5.06 \times 10^8$~$\psi'$ events have been used.
For these, known decay modes are modelled with \textsc{evtgen}\cite{Ping:2008zz} using branching fractions taken from the Particle Data Group, whereas unknown decay modes are generated following the \textsc{lundcharm} model~\cite{Chen:2000tv}. 
The main background channels are found to be $\Psi\to\Delta^{+} \bar{\Delta}^{-}$ and $\Psi\to\gamma \eta_{c}, \eta_{c} \to \Sigma^{+} \bar{\Sigma}^{-}$ using the tool described in Ref.~\cite{Zhou:2020ksj}. Here, the latter channel already only constitutes about $0.07~\%$ of the signal strength and can thus be neglected. 

\begin{figure}[htbp]
\includegraphics[width=0.4\textwidth]{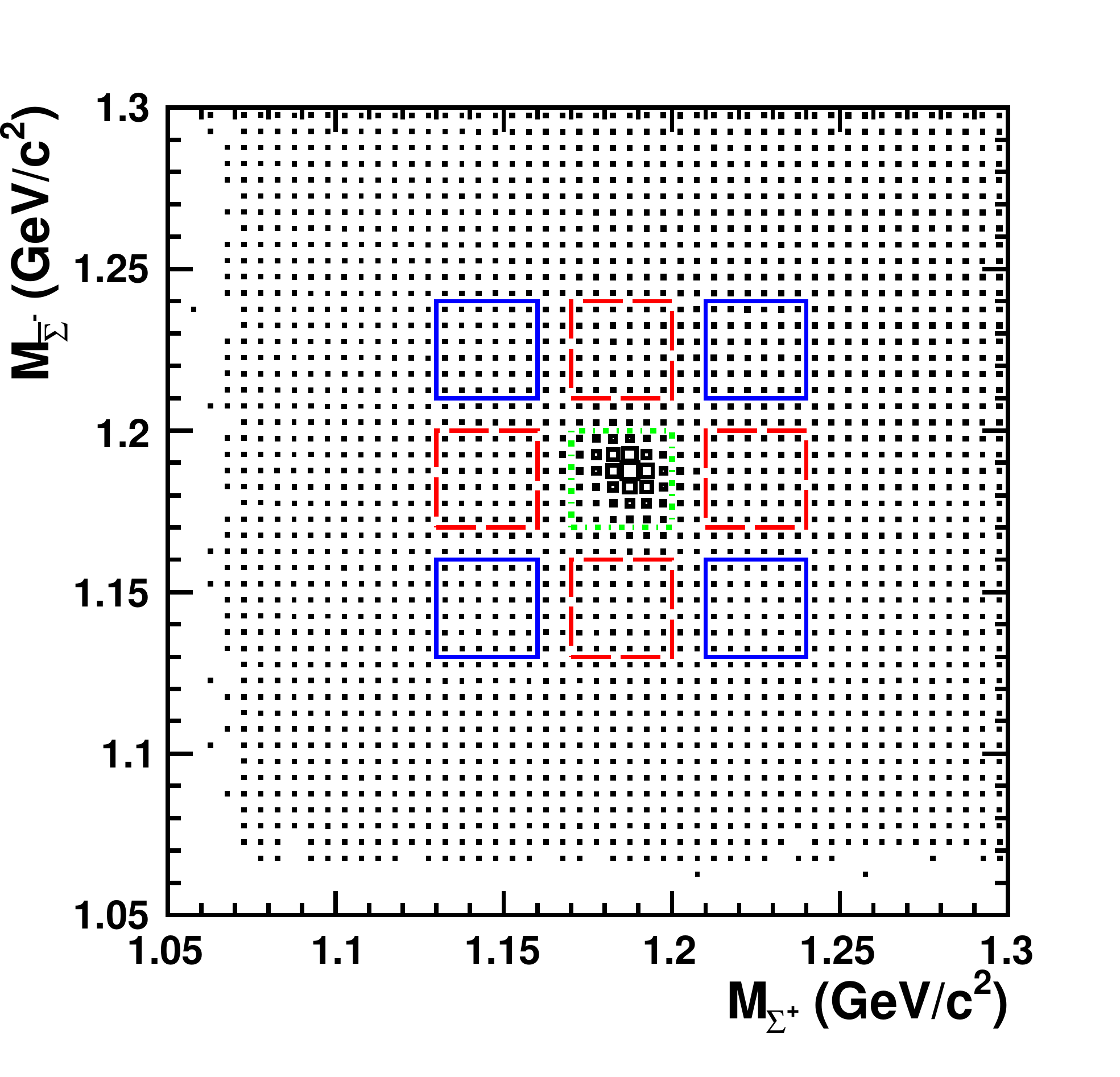}
\caption{(Color online) Distribution of the invariant mass of the $\bar{\Sigma}^{-}$ candidate as a function of the invariant mass of the $\Sigma^{+}$ candidate, showing the 
signal (green box) and sideband regions A and B (red and blue boxes, respectively).}
\label{fig:background}
\end{figure}

\begin{figure}[htbp]
\includegraphics[width=0.2\textwidth,angle=0]{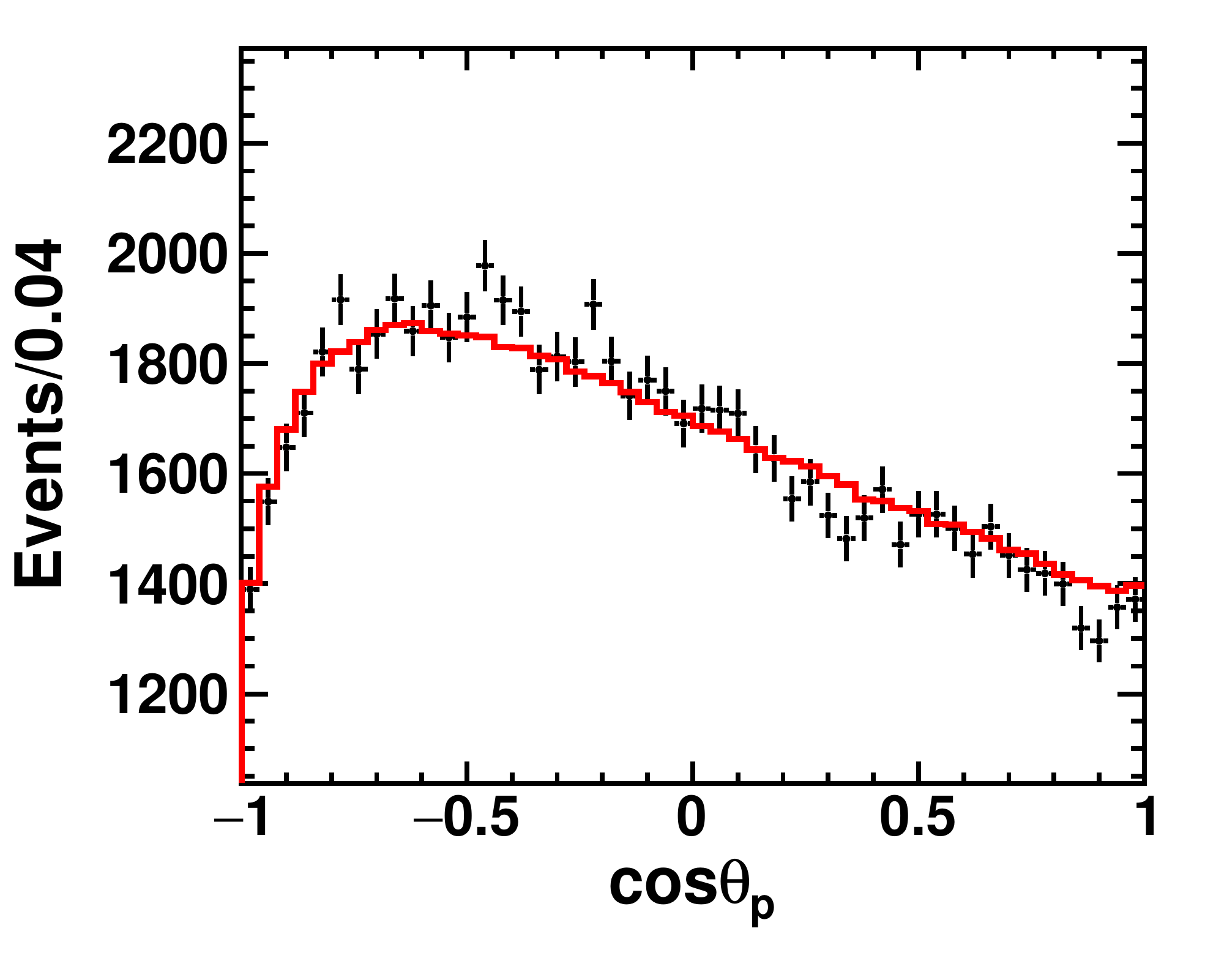}
\includegraphics[width=0.2\textwidth,angle=0]{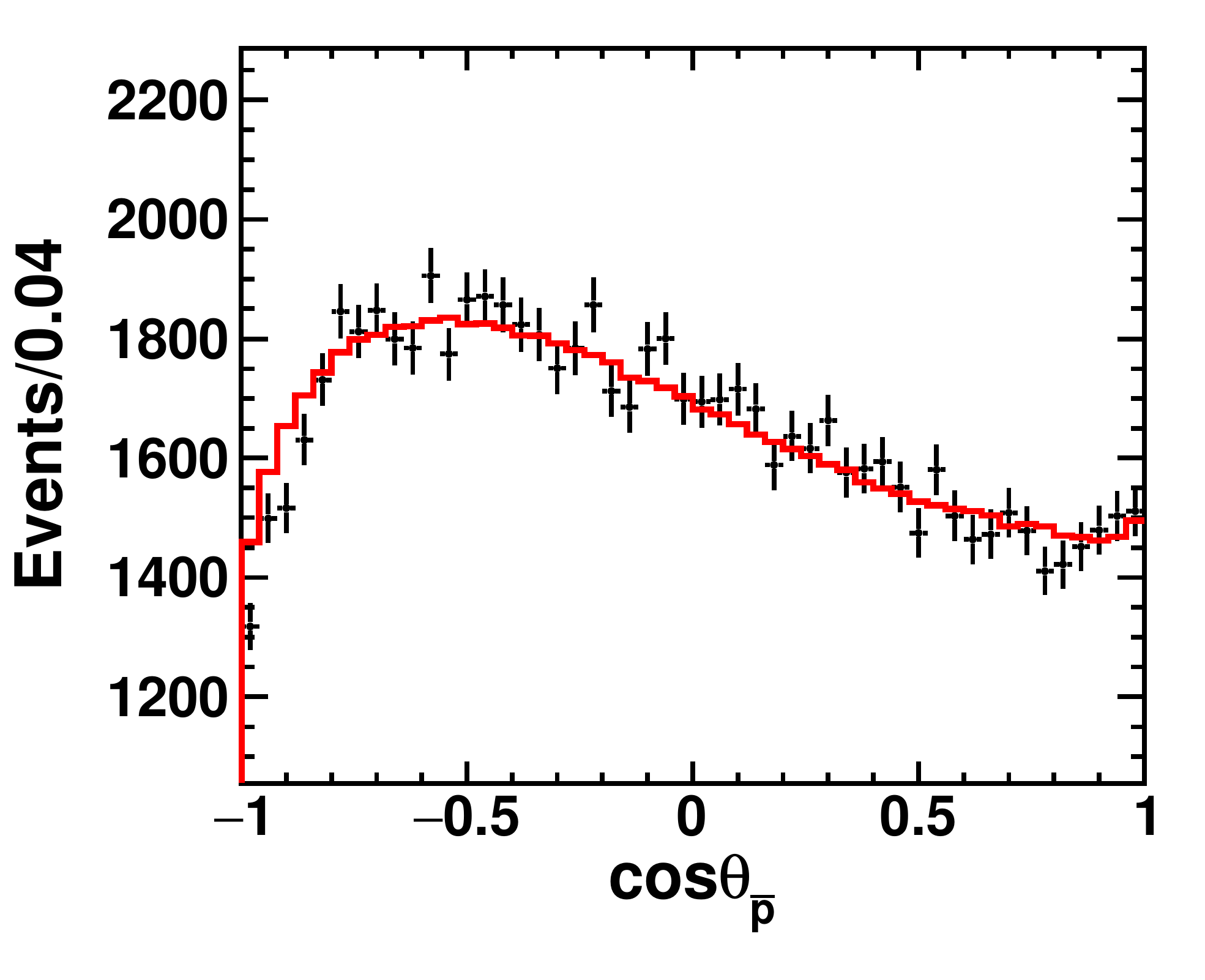}\\
\includegraphics[width=0.2\textwidth,angle=0]{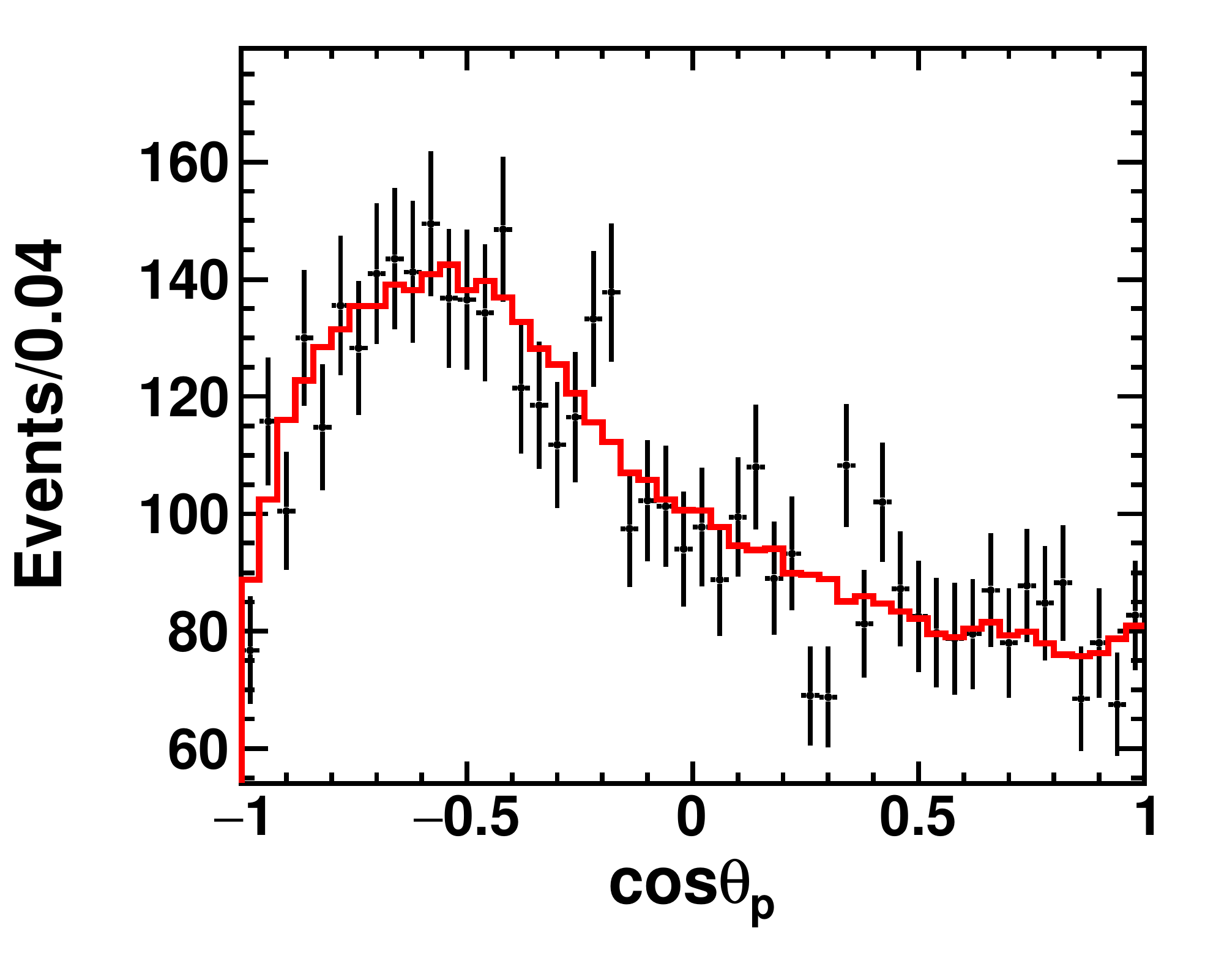}
\includegraphics[width=0.2\textwidth,angle=0]{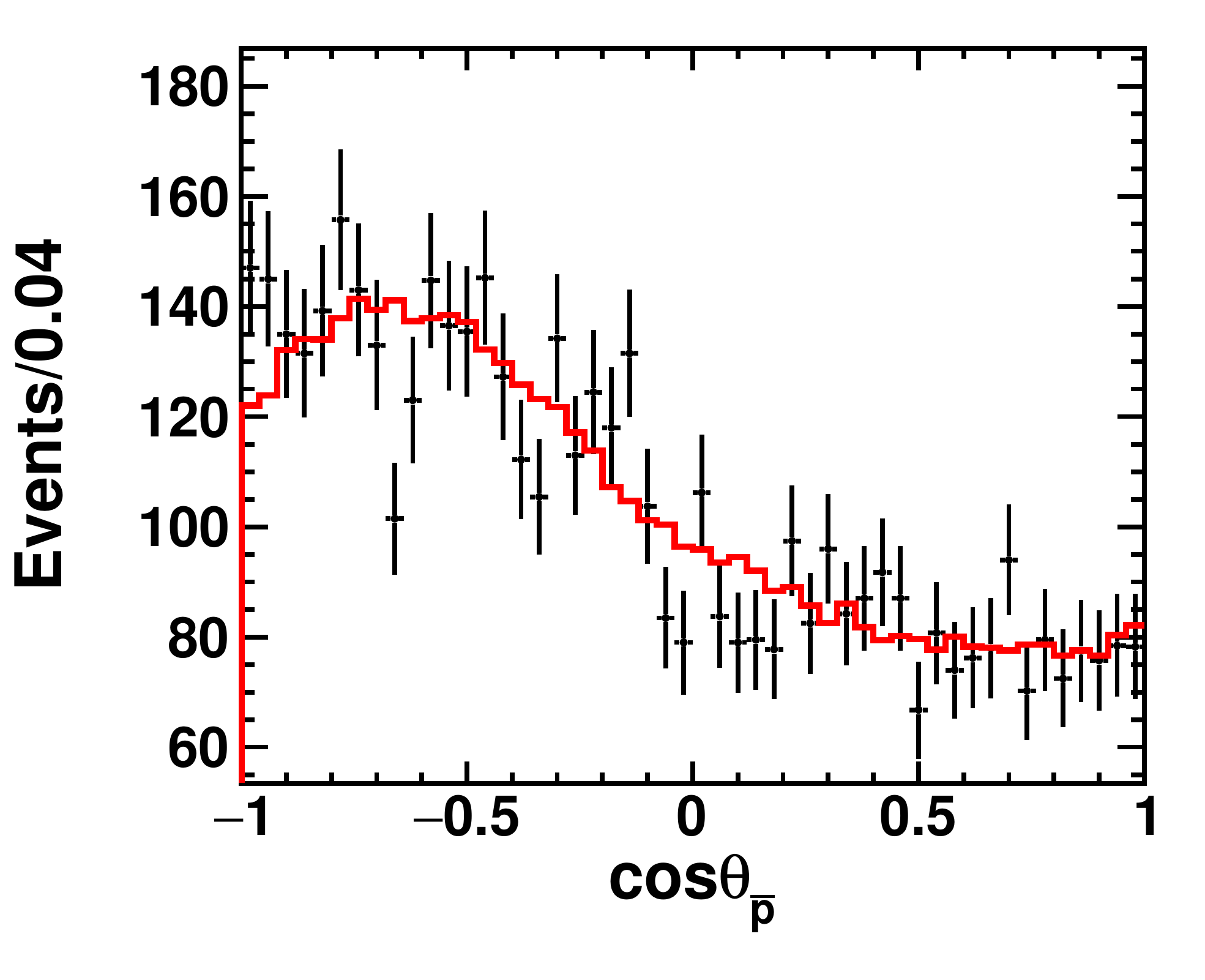}
\caption{(Color online) Comparison between data and MC simulated events weighted by fit results in $\cos\theta_{p}$ (top-left), $\cos\theta_{\bar{p}}$ (top-right) for $J/\psi$ decay and $\cos\theta_{p}$ (bottom-left), $\cos\theta_{\bar{p}}$ (bottom-right) for $\psi'$ decay. The black points with error bars are experimental data and red solid lines represents the MC-determined shapes taking into account the fit results.}
\label{fig:cos_p}
\end{figure}

To estimate the amount of non-$\Sigma^{+} \bar{\Sigma}^{-}$ events in data, a two-dimensional sideband method is used to quantify the background contribution.
The signal region is defined as $1.17~\gevcc < M_{p\pi^{0} / \bar{p}\pi^{0}} < 1.2~\gevcc$ and the lower and upper sideband regions are defined as $1.13~\gevcc < M_{p\pi^{0} / \bar{p}\pi^{0}} < 1.16~\gevcc$ and $1.21~\gevcc < M_{p\pi^{0} / \bar{p}\pi^{0}} < 1.24~\gevcc$, respectively.
The sideband regions are shown in Fig.~\ref{fig:background}. We discriminate between two different types of sideband contributions. Regions A, indicated with red dashed lines in Fig.~\ref{fig:background}, designate those events where one of the $p\pi^0$ or $\bar{p}\pi^0$ combinations lies in the signal region while the other one does not, whereas regions B, indicated as blue solid lines, designate events where both $p\pi^0$ and $\bar{p}\pi^0$ fall into the respective sideband.
The number of background events $N_{bg}$ is then determined by $N_{bg} = 0.5N_A - 0.25N_B$, where $N_A$ and $N_B$ are the sum of all events in the regions A and B, respectively.
From this method, the background levels in the signal region (green dotted box in Fig.~\ref{fig:background}) are found to be $5\%$ for $J/\psi \rightarrow \Sigma^{+}\bar{\Sigma}^{-}$ and $1\%$ for $\psi' \rightarrow \Sigma^{+}\bar{\Sigma}^{-}$. The final event samples in the signal region are determined to be 87815 events for the $J/\psi\to\Sigma^{+}\bar{\Sigma}^{-}$ decay and 5327 events for the $\psi'\to\Sigma^{+}\bar{\Sigma}^{-}$ decay. 
An unbinned maximum likelihood fit is performed in the five angular dimensions ${\boldsymbol{\xi}}$, simultaneously fitting both the $J/\psi \rightarrow \Sigma^{+}\bar{\Sigma}^{-}$ and $\psi' \rightarrow \Sigma^{+}\bar{\Sigma}^{-}$ data in order to determine the parameters ${\boldsymbol{\Omega}}=\{\alpha_{J/\psi}, \alpha_{\psi'}, \Delta\Phi_{J/\psi}, \Delta\Phi_{\psi'}, \alpha_{0}, \bar{\alpha}_{0}\}$.
In the fit, the joint likelihood function is defined as
\begin{equation*}
\mathscr{L} = \prod_{i=1}^n Prob({\boldsymbol{\xi}}_i, {\boldsymbol{\Omega}}) = \prod_{i=1}^n \frac{{\cal{W}({\boldsymbol{\xi_i}}, {\boldsymbol{\Omega}})}}{\cal{N}},  
 \end{equation*}
\noindent where $n$ is the number of events and $Prob({\boldsymbol{\xi}}_i)$ is the probability to produce event $i$ based on the measured observables ${\boldsymbol{\xi}}$ and the set of parameters $\boldsymbol{\Omega}$. The normalization factor $\mathcal{N}=\frac{1}{N_\mathrm{MC}}\cdot\sum_{j=1}^{N_\mathrm{MC}}\mathcal{W}_j^\mathrm{MC}$ is given by the sum of the corresponding amplitude $\cal{W}$ using simulated events evenly distributed in phase space. 
In the normalization factor, the detection efficiency is included and possible differences between real data and MC simulations have been taken into account.
Instead of the likelihood function $\mathscr{L}$, the negative of the logarithm of $\mathscr{L}$ is minimized using the MINUIT package given in the CERN library~\cite{James:1975dr, James:1994vla}.
The objective function is defined as
\begin{equation*}
\mathit{S} = -\mathrm{ln}\mathscr{L}_{data} + \mathrm{ln}\mathscr{L}_{bg},
\end{equation*}
where $\mathscr{L}_{data}$ is the likelihood function of events selected in the signal region, and $\mathscr{L}_{bg}$ is the likelihood function of background events given by the sideband regions.
The numerical fit results are summarized in Table~\ref{table:sum_decay}. The first uncertainty given is always statistical and the second one is systematic. In Fig.~\ref{fig:cos_p}, the fit results of asymmetry parameters are illustrated using $\cos\theta_{p}$ and $\cos\theta_{\bar{p}}$ projections, which follow $dN/d\Omega = \frac{1}{4\pi}( 1 + \alpha_{0}\mathbf{P}_{\Sigma^{+}} \cdot \mathbf{\hat{p}})$ distribution.

\begin{figure*}[htbp]
\includegraphics[width=0.4\textwidth,angle=0]{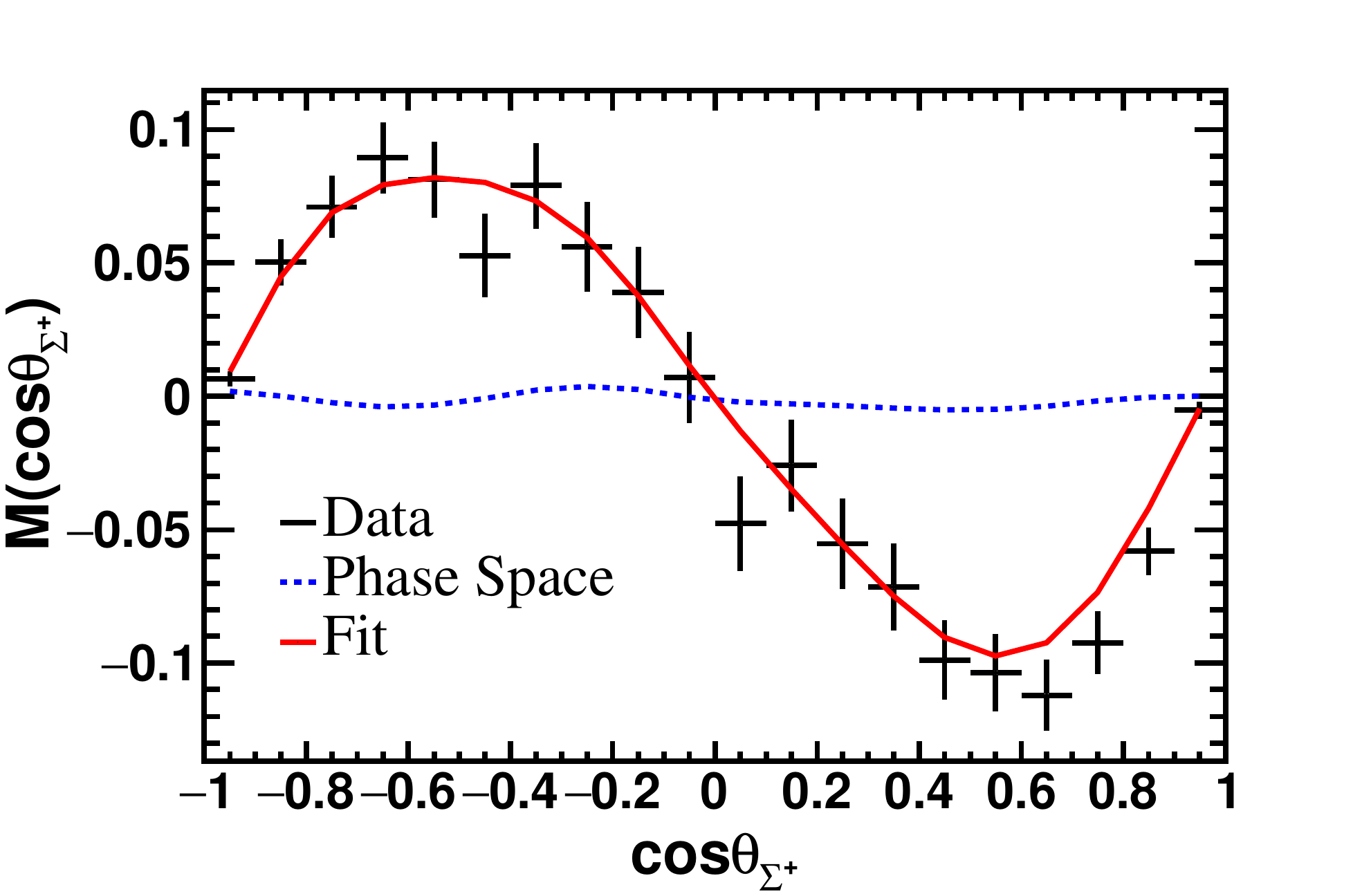}
\includegraphics[width=0.4\textwidth,angle=0]{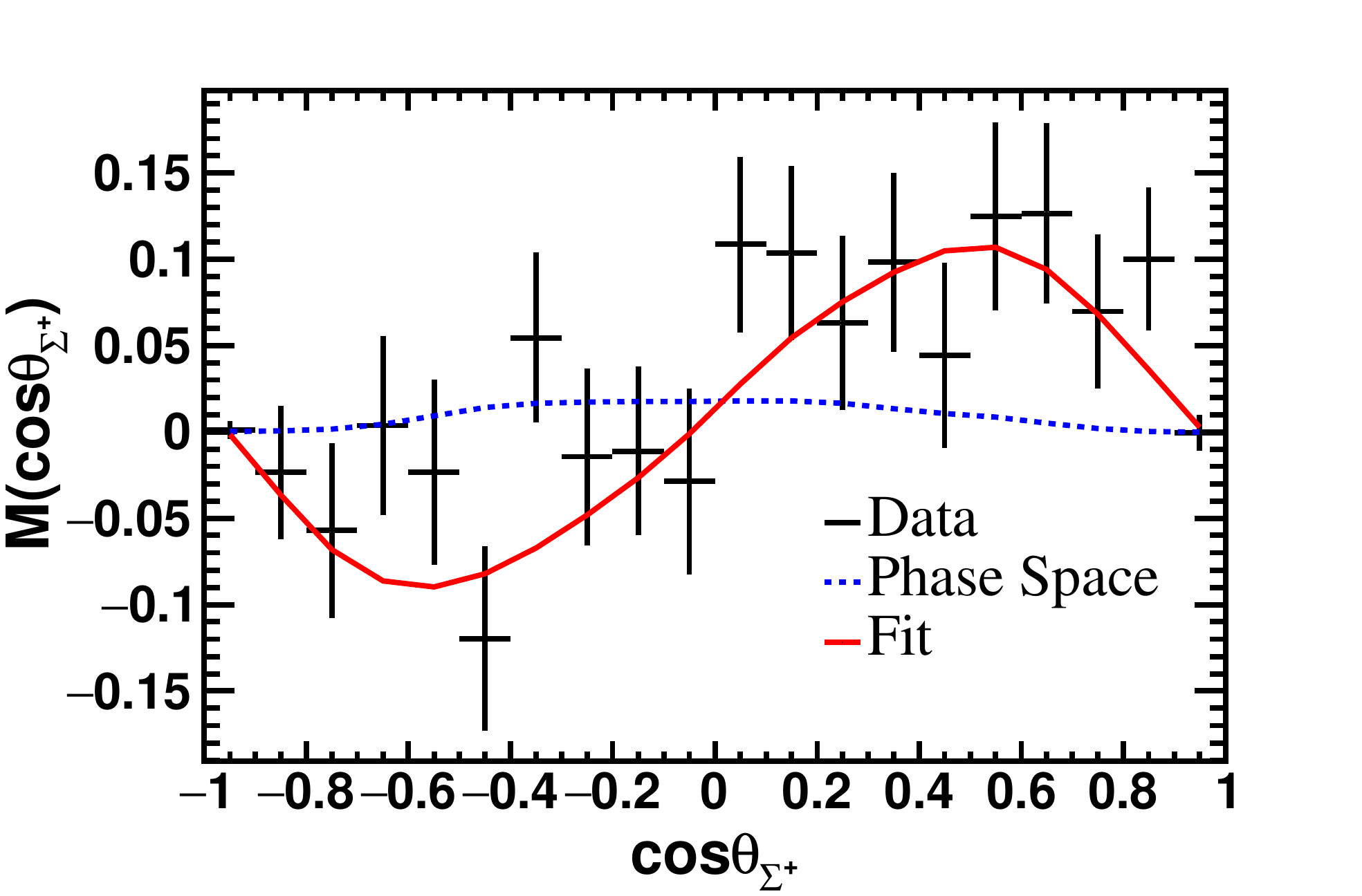}
\caption{(Color online) The moments $M(\cos\theta_{\Sigma^{+}})$ for data that is not corrected for acceptance and reconstruction efficiency as function of $\cos\theta_{\Sigma^{+}}$ for the decays $J/\psi\to\Sigma^{+}\bar{\Sigma}^{-}$ (left) and $\psi' \rightarrow \Sigma^{+}\bar{\Sigma}^{-}$ (right). The black points with error bars are experimental data, the red solid lines are the fit results and the blue dashed line represents
the distribution without polarization from the simulated events evenly distributed in phase space. }
\label{fig:data_fitting}
\end{figure*}

\begin{table}[htbp]
\caption{\label{table:sum_decay} Values and uncertainties of the fit parameters extracted in this work. }
\begin{ruledtabular}
\begin{tabular}{c c}
Parameter  & Measured value \\
\hline
$\alpha_{J/\psi}$                         &$-$0.508 $\pm$ 0.006 $\pm$ 0.004  \\
$\Delta\Phi_{J/\psi}$                  & $-$0.270 $\pm$ 0.012 $\pm$ 0.009 \\
$\alpha_{\psi'}$                  &0.682 $\pm$ 0.03  $\pm$  0.011 \\
$\Delta\Phi_{\psi'}$           &0.379 $\pm$ 0.07   $\pm$ 0.014 \\
$\alpha_{0}$             &$-$0.998 $\pm$ 0.037 $\pm$ 0.009  \\
$\bar{\alpha}_{0}$      &0.990 $\pm$ 0.037  $\pm$ 0.011  \\
\end{tabular}
\end{ruledtabular}
\end{table}
The spin polarization of the $\Sigma$ baryons is observed for both the $J/\psi$ and the $\psi'$ datasets. The relative phase between the psionic electric and magnetic form factors is determined to be $\Delta\Phi_{J/\psi}=-0.270 \pm 0.012$ and $\Delta\Phi_{\psi'}=0.379 \pm 0.07 $ for the $J/\psi\to \Sigma^{+}\bar{\Sigma}^{-}$ and $\psi' \to \Sigma^{+}\bar{\Sigma}^{-}$ decay, respectively, which differs from zero with a significance of more than 20 $\sigma$ in case of the $J/\psi$ data and with a significance of $5.5 \sigma$ for the $\psi'$ data, including systematic uncertainties.
The two values determined at the $J/\psi$ and $\psi'$ resonances differ in size and also have oppsite sign.
The polarization of the $\Sigma$ baryons is clear visible in the data, as shown in Fig.~\ref{fig:data_fitting}, where the moment $M(\cos\theta_{\Sigma^{+}})$ is displayed for the data divided into 20 $\cos\theta_{\Sigma^{+}}$ bins in comparison to a MC sample evenly distributed in phase space and the solution of the fit performed in this work. The moment is given by
\begin{equation*}
 M(\cos\theta_{\Sigma^{+}}) = (m/N)\sum_i^{N(\cos\theta_{\Sigma^{+}})}(\sin\theta_{p}^{i} \cos\phi_{p}^{i} - \sin\theta^{i}_{\bar{p}} \cos\phi_{\bar{p}}^{i}). 
\end{equation*}
Here, $m=20$ is the number of bins, $N$ is the total number of events in the data sample and $N(\cos\theta_{\Sigma^{+}})$ is the number of events in the $\cos\theta_{\Sigma^{+}}$ bin. Assuming CP-conservation $\alpha_{0} = - \bar{\alpha}_{0}$, the expected angular dependence of the moment from Eq.~(\ref{eq:anglW}) is $\frac{dM}{d\cos\theta_{\Sigma^{+}}}\sim\sqrt{1-\alpha_{\psi}^2}\alpha_{0} \sin\Delta\Phi \cos\theta_{\Sigma^{+}} \sin\theta_{\Sigma^{+}}$ in case of data corrected for the acceptance and reconstruction efficiency. The red line in Fig.~\ref{fig:data_fitting} follows this expectation but additionally takes acceptance and reconstruction efficiency into account.

As $\Delta\Phi$ is non-zero, a simultaneous measurement of $\alpha_{0}$ and $\bar{\alpha}_{0}$ is possible and performed, as shown in Table~\ref{table:sum_decay}. 
From the asymmetry parameters $\alpha_{0}$ and $\bar{\alpha}_{0}$, the CP-odd observable $A_{\rm CP,\Sigma} = (\alpha_0 + \bar{\alpha}_0)/(\alpha_0 - \bar{\alpha}_0) = -0.004\pm0.037\pm0.010$ 
is extracted for the first time. It is found to be consistent with the standard model prediction.
The average decay asymmetry $(\alpha_{0} - \bar{\alpha}_{0})/2$ is calculated to be $-0.994\pm0.004\pm0.002$, representing a significant improvement in precision compared to earlier measurements.

A summary of the systematic uncertainties that have been considered in this work are listed in Table~\ref{table:sys_decay}.
Sources under consideration include a possible bias in the fit method, the choice of mass region for the signal, the background estimation method, helix parameter corrections and efficiency differences between data and MC simulations. The individual uncertainties are assumed to be uncorrelated and are therefore added in quadrature.
To validate the reliability of the fit results, a set of 100 toy samples is simulated. In these samples, the differential cross section is based on Eq.~(\ref{eq:anglW}), and the decay parameters determined in this study, listed in Table \ref{table:sum_decay}, are used as input parameters. The number of events in each toy sample is the same as for the data sample. We compare the average output values with the input values for all fit parameters. Differences between input and average output are taken as the systematic uncertainties caused by the fitting method.
In addition, the size of the signal mass window is changed by $\pm 5~\mev$. The fit is repeated and the differences between the new values and the nominal values are taken as the systematic uncertainties of the parameters resulting from the choice of signal mass window.
In this work, the sideband regions in the $\Sigma^{+}\to p\pi^0$ and $\bar{\Sigma}^{-}\to \bar{p}\pi^0$ invariant masses were used to estimate the amount of background events in the signal region. Changing the sideband regions from [1.13, 1.16] $\gevcc$ and [1.21, 1.24] $\gevcc$ to [1.145, 1.16] $\gevcc$ and [1.21, 1.225] $\gevcc$, the background estimation and the fit are repeated and
the differences between the new and the nominal fit results are taken as the systematic uncertainties on the fit parameters caused by the choice of sideband regions.
For the nominal result, we are using the track correction for the helix parameters mentioned in Ref.~\cite{Ablikim:2012pg}. We repeat the full fit procedure using a MC sample without this track correction and take the difference between the two fit results as a systematic uncertainty caused by the track correction.
The uncertainties due to potential efficiency differences between data and simulations of charged-particle tracking and PID have been investigated with $J/\psi \rightarrow p \bar{p} \pi^{+} \pi^{-}$ control samples, and those due to neutral $\pi^{0}$ reconstruction are estimated from $J/\psi \rightarrow \pi^{+} \pi^{-} \pi^{0}$ control samples. Using these control samples, we determine corrections to the MC simulations and take the differences between fit results with and without tracking, PID and $\pi^{0}$ reconstruction efficiency corrections as the systematic uncertainties.

\begin{table}[htbp]
\caption{Summary of the systematic uncertainties on the resulting fit parameters. }
\begin{center}
\begin{tabular}{c | c c c c c c}
\hline
\hline
Source &$\alpha_{J/\psi}$   &$\Delta\Phi_{J/\psi}$ &$\alpha_{\psi'}$  &$\Delta\Phi_{\psi'}$  &$\alpha_{0}$   &$\bar{\alpha}_{0}$ \\
\hline
Fit method                        &0.002       &0.004       &0.005     &0.011    &0.007  &0.008\\
Signal window                &0.002         &0.006       &0.008.   &0.007     &0.003  &0.005\\
Background                    &0.002       &0.005       &0.003     &0.002    &0.002   &0.001\\
Track correction               &0.000       &0.001       &0.003     &0.000    &0.004   &0.005\\
Eff. correction                   &0.000       &0.001        &0.003    &0.000    &0.000  &0.001\\
\hline
Total                                   &0.004       &0.009       &0.011     &0.014    &0.009    &0.011\\

\hline
\hline
\end{tabular}
\end{center}
\label{table:sys_decay}
\end{table}%

In conclusion, based on the samples of 1310.6 $\times 10^6$ $J/\psi$ and 448.1 $\times 10^{6}$ $\psi'$ events collected with the BESIII detector, 
the decay parameters of the decays $J/\psi \rightarrow \Sigma^{+} \bar{\Sigma}^{-}$ and $\psi' \rightarrow \Sigma^{+} \bar{\Sigma}^{-}$, $\alpha_{J/\psi}$ and $\alpha_{\psi'}$, are measured for the first time. The numerical fit results are given in Table~\ref{table:sum_decay}.
Here, $\alpha_{J/\psi}$ is determined to be negative, which has the same sign as observations made in the decays $J/\psi \to \Sigma^{0}\bar{\Sigma^{0}}$, $J/\psi \to \Sigma(1385)^{-} \bar{\Sigma}(1835)^{+}$ and $J/\psi \to \Sigma(1385)^{+} \bar{\Sigma}(1835)^{-}$~\cite{Ablikim:2016iym}.

The relative phases $\Delta\Phi_{J/\psi}$ and $\Delta\Phi_{\psi'}$ are determined simultaneously
and for the first time for both reactions $J/\psi \rightarrow \Sigma^{+} \bar{\Sigma}^{-}$ and $\psi' \rightarrow \Sigma^{+} \bar{\Sigma}^{-}$. This also marks the first determination of the relative phase for a $\psi'$ decay into a pair of baryons.
Since $\Delta\Phi$ is found to be non-zero for both decays, the decay asymmetry parameters $\alpha_{0}$ and $\bar{\alpha}_{0}$ are determined simultaneously. 
While the value of $\alpha_{0}$ determined in this work is consistent with the PDG average at significantly improved precision, $\bar{\alpha}_{0}$ is measured for the first time. The value of $A_{\rm CP, \Sigma}$ is found to be consistent with CP-conservation and is in agreement with the Standard Model prediction within present uncertainties~\cite{Tandean:2002vy}.
\\

\begin{acknowledgments}
The BESIII collaboration thanks the staff of BEPCII and the IHEP computing center for their strong support. This work is supported in part by National Key Basic Research Program of China under Contract No. 2015CB856700; National Natural Science Foundation of China (NSFC) under Contracts Nos. 11625523, 11635010, 11735014, 11805037, 11822506, 11835012, 11935015, 11935016, 11935018, 11961141012; the Chinese Academy of Sciences (CAS) Large-Scale Scientific Facility Program; Joint Large-Scale Scientific Facility Funds of the NSFC and CAS under Contracts Nos. U1732102, U1732263, U1832103, U1832121, U1832207; CAS Key Research Program of Frontier Sciences under Contracts Nos. QYZDJ-SSW-SLH003, QYZDJ-SSW-SLH040; 100 Talents Program of CAS; INPAC and Shanghai Key Laboratory for Particle Physics and Cosmology; ERC under Contract No. 758462; German Research Foundation DFG under Contracts Nos. Collaborative Research Center CRC 1044, FOR 2359; Istituto Nazionale di Fisica Nucleare, Italy; Ministry of Development of Turkey under Contract No. DPT2006K-120470; National Science and Technology fund; STFC (United Kingdom); The Knut and Alice Wallenberg Foundation (Sweden) under Contract No. 2016.0157; The Royal Society, UK under Contracts Nos. DH140054, DH160214; The Swedish Research Council; Olle Engkvist Foundation under Contract No 200-0605; U. S. Department of Energy under Contracts Nos. DE-FG02-05ER41374, DE-SC-0012069.
\end{acknowledgments}

\end{document}